# picoArray technology: the tool's story

Andrew Duller, Daniel Towner, Gajinder Panesar, Alan Gray, Will Robbins
picoChip Designs Ltd., Riverside Buildings,
108 Walcot Street, Bath BA1 5BG, UK
andy.duller@picochip.com

**Abstract**

*This paper briefly describes the picoArray<sup>TM</sup> architecture, and in particular the deterministic internal communication fabric. The methods that have been developed for debugging and verifying systems using devices from the picoArray family are explained. In order to maximize the computational ability of these devices, hardware debugging support has been kept to a minimum and the methods and tools developed to take this into account.*

## 1. Introduction

The wireless communications field is experiencing a period of major expansion. This is happening all over the world and is not dominated by any one region in particular. In a field in which different standards are fixed for different regions of the world, where standards are in a state of flux or even where no standards exist, it is very costly to enter the market with a custom ASIC solution. What is required is a scalable programmable solution, which can cater for most, if not all, of these areas. To this end the picoArray and a rich toolset have been created.

The picoArray is a tiled processor architecture in which hundreds of processors are connected together using a deterministic interconnect [4, 3]. The level of parallelism is relatively fine grained with each processor having its own small amount of data and instruction memory. Each processor runs a single process in its own memory space and uses "signals" to synchronise and communicate. Multiple picoArray devices may be connected together to form systems containing thousands of processors using peripherals which effectively extend the on-chip bus structure.

In order to provide a massively parallel, scalable solution that is commercially viable, it has been necessary to re-think methods of debug and verification in the following areas:

**Scale** Depending upon the target, systems solutions may require moderate or massive computational power. To address this, many picoArray devices may be connected together[1], creating systems containing thousands of processors.

**Reduced non-essential hardware** Specialised hardware for anything other than computation must be justified and the emphasis should be on system-wide rather than processor-centric debug hardware. Therefore conventional support, such as register or memory trace mechanisms, is not as useful as in a uni-processor system. In the picoArray, hardware support for debug has been kept to a minimum in order to allow more processors to be fitted onto a single device.

**Environment** The picoArray devices are designed for use in embedded environments where there are relatively few inputs and outputs. In such environments, the bandwidth available for debugging traffic is limited. Relatively light weight access is provided to the picoArray via a JTAG interface and the external processor interface.

**Communication and synchronisation** Many parallel systems use special purpose libraries (e.g., Posix threads [2], Message Passing Interface [8]) or language support mechanisms (e.g., Java threads [5]) to handle communication and synchronisation of parallel processes. This is very difficult to justify in an embedded system where memory cost is an area of concern. The picoArray uses a deterministic interconnect fabric called the picoBus. This behaves like a handshaken FIFO between processes, and is used for communication and synchronisation. No run-time arbitration of the picoBus is necessary, leading to simpler hardware and removing a possible source of bugs.

Conventional debuggers are usually designed for a single processor, and normally require a reasonable amount of hardware support. Even debuggers from vendors who claim to support multi-processor systems will require hardware support. More importantly, however, conventional debuggers will not scale to systems consisting of hundreds or thousands of processors.

---

1 Boards with up to 16 devices have been produced.

The approach to debug and verification has been to exploit the programming paradigm provided by the picoArray [4, 3] and to create a rich set of tools to aid system-wide debug and verification. The approach can be divided into two main aspects: language features and software tools.

The picoArray tools support an input language which is a combination of VHDL [1], ANSI/ISO C and assembly language. Individual processes are written in C and assembler, while structural VHDL is used to describe how processes are connected together using signals. Signals are strongly typed, and have specified bandwidths. They may be synchronous or asynchronous, point-to-point or point-to-multipoint. Processes are statically created by describing their number and type in the source files — no runtime creation of processes is possible. Thus, after a system has been compiled, the complete set of processes and their connectivity is known, which ensures that the system will behave deterministically.

The most common software tool for debugging is the symbolic debugger, which allows the programmer's original source code to be displayed, along with the contents of source variables, on a per-process basis. The use of symbolic debugging is commonplace, so will not be considered further in this paper. The additional mechanisms that have been developed are the following:

**design browser** This allows both static and dynamic analysis of a user's design. The overall structure of a system can be graphically displayed in a number of ways. This allows the user to verify that the overall system has been connected in the way that was intended and to visualise and navigate through a complex design which may have been coded by many people. In addition, dynamic analysis can be performed and it can be used to visualise problems such as those associated with data throughput and deadlock.

**simulation** The cycle accurate simulator allows all of the processor's internal state to be viewed, including aspects that the real hardware does not allow access to. Typically the user would start their development here and then migrate to hardware.

**scripting** The debugger can be programmed using Tcl/Tk. This allows the user to build on the basic system provided by the standard debugger.

**probes** These are processes which can be inserted into a user's design during debugging, to enable complex real-time analysis.

**file I/O** Data can be streamed in to and out of a system using files.

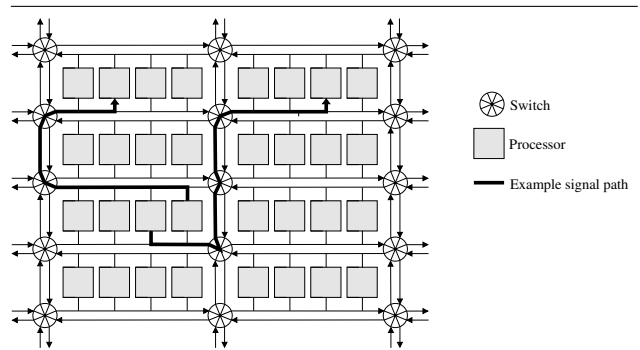

**Figure 1. picoArray Interconnect**

The remainder of this paper gives an overview of picoArray devices, a brief description of each debug mechanism, and then shows how the debug and verification tools might be used over the lifetime of a design, from initial component to integrated system.

## 2. The picoArray Concept

### 2.1. The picoArray Architecture

The latest device - PC102 is based around the picoArray tiled processor architecture in which over 300 processors (3-way VLIW, Harvard architecture with local memory), and 14 co-processors (Function Accelerator Units or FAUs) are interconnected by a 32-bit picoBus and programmable switches.

The term Array Element (AE) is used to describe either processors or co-processors (i.e. there are 322 AEs in the array). There are three processor variants which share the same basic structure: Standard AE (STAN), Control AE (CTRL) and Memory AE (MEM). The memory configuration and number of communications ports varies between AE types.

### 2.2. Inter-processor Communications

Within the picoArray core, AEs are organised in a two dimensional grid, and communicate over a network of 32-bit buses (the picoBus) and programmable bus switches. AEs are connected to the picoBus by ports. The ports act as nodes on the picoBus and provide a simple interface to the bus based on *put* and *get* instructions.

The inter-processor communication protocol is based on a time division multiplexing (TDM) scheme, where data transfers between processor ports occur during time slots scheduled automatically by the tools and controlled using the bus switches. The bus switch programming and the scheduling of data transfers is fixed at compile time and requires no run-time arbitration. Figure 1 shows an example in which the switches have been set to form two different signals between processors. Signals may be point-to-point or point-to-multi-point. In the latter case, the data transfer will not take place until all the processor ports involved in the transfer are ready. The total internal data bandwidth for the signals is 3.3 Tera-bits per second [2].

The default signal transfer mode is synchronous; data is not transfered until both the sender and receiver ports are ready for the transfer. If either is ready before the other then the transfer will be retried during the next available time slot. If, during a *put* instruction, no buffer space is available then the processor will sleep (hence reducing power consumption) until space becomes available. In the same way, if during a *get* instruction there is no data available in the buffers then the processor will also sleep. This protocol ensures that no data can be lost.

There is also an asynchronous signal mode where transfer of data is not handshaken and in consequence data can be lost by being overwritten in the buffers without being read.

### 2.3. External Communications

The picoArray has three methods of external communications. They are:

- External Processor Interface (EPI),
- Inter-picoArray Interface (IPI),
- Asynchronous Data Interface (ADI).

These can all be connected to the picoBus and can be accessed using signals. The EPI can be used to configure picoArray devices and can be used by debugging tools for input and output of (2.5 Giga-bits per second) information.

The IPI is used to connect picoArray devices together and can be viewed as a way of extending the picoBus across devices.

The ADI is used for exchanging data with high bandwidth (5 Giga-bits per second) external asynchronous data streams.

Each device has a single EPI and four interfaces each of which can be configured as either an IPI or an ADI.

## 3. picoArray Debug and Analysis

### 3.1. Language Features

The language features aid verification and integration through three main features: strong type checking, fixed process creation, and bandwidth allocation.

---

2    322 processors x 2 buses x 32-bits x 160MHz clock

Strong type checking is used to ensure that whenever data is communicated from one process to another, the data will be interpreted by both producer and consumer in the same way. Types are selected from a library of built-in types, or by the user defining their own types. Types used in communication are limited to 32-bits, which is the maximum size which may be transferred in a single communication over the picoBus. At the structural level, processes will be defined with ports of specific types, and they will be connected with signals which must match the port types. Within a process, any data which is *put* or *get* from a port must be of the correct type. For processes written in C, this is achieved by synthesising the available types using C encoding rules (e.g., using typedef's, union's, and struct's), and hence tying in to the C compiler's type system. Thus, end-to-end communication of data can only occur when all processes and signals agree on the type format. This facilitates integration of independently developed components since any discrepancies in type formats will be detected at compile time.

The structural VHDL used to define a system requires the number of processes, and their interconnections, to be fixed at compile time. During compilation, the tools will allocate each process to its own processor and schedule the signals connecting the processes on to the picoBus interconnection fabric. Because of this compile-time scheduling, non-deterministic runtime effects such as process scheduling or bus contention have been eliminated. This makes it easier to integrate systems. If problems are found, it also makes the reproduction of the problems, their debugging and the verification of their fixes easier.

In addition to specifying fixed signals connecting processes, the signals are also allocated bandwidth. This is achieved using a language notation which allows the frequency of communication over the signal to be specified. Processes requiring high signal bandwidths will use high frequencies (e.g., every 4 cycles), while processes requiring low bandwidth will use low frequencies (e.g., every 1024 cycles).

### 3.2. Design Browser

The design browser is a tool which allows the user's logical design to be viewed graphically and can be used both during simulation and when executing a design on hardware. The following different graphical views are possible:

- hierarchical,
- flat within a given scope,
- as the strongly connected components (SCC).

The hierarchical view mirrors the structural hierarchy that was created by the user and allows each level of this hierarchy to be explored. An example of this is shown in figure 2.

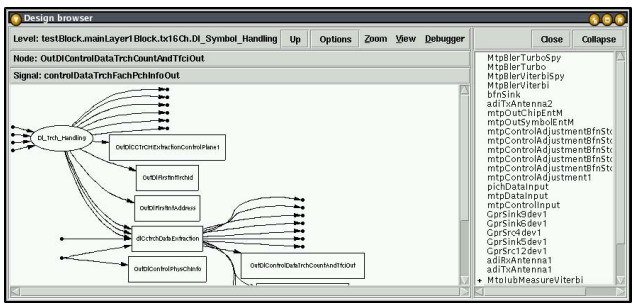

**Figure 2. Design browser hierarchical display**

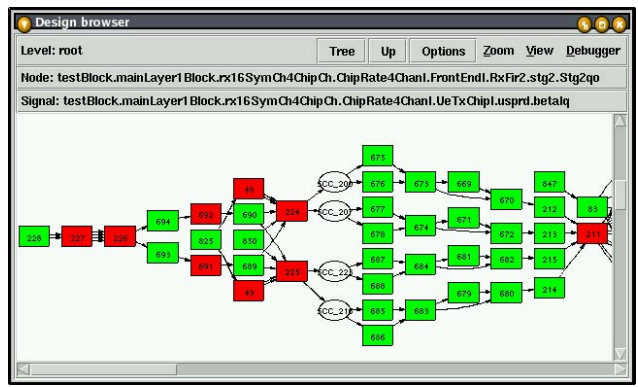

**Figure 3. Design browser strongly connected component display**

There are times when the user wishes to see more of a design than is permitted by the hierarchy display, and the "flat" display provides this. If displayed from the root of the design the entire design is displayed at once. In addition, by displaying from a scope other than the root, subtrees of the design can be viewed. The major difference between this and the hierarchy display is that from a given scope all of the leaf instances are displayed at once.

The final view comes from thinking of a design as a directed graph and then producing a single level of hierarchy by identifying the strongly connected components (SCC). Each of these components can be viewed on their own. An example of this is shown in figure 3. The importance of the SCC view is that from the root level the graph becomes acyclic (a directed acyclic graph) and this gives advantages when trying to debug a system which has deadlock, livelock or data throughput problems, because it separates out the parts of the design that contain feedback from those that are simply pipeline processing.

In addition to these static features the design browser can provide dynamic information about the each instance in a design, for example whether it is processing or waiting for a communications operation. An example of this display is shown in figure 3 where the boxes are coloured green for processing and red for waiting on communications.

### 3.3. Simulation

The cycle accurate simulation system allows users to build, test and verify their entire design before moving to the hardware. The user is able to extract the state of the system (on a cycle-by-cycle basis) in order to check against the behaviour on hardware. Importantly, the same simulation system was used to provide a "golden reference" during the design and verification of the PC101 and PC102 chips.

The same source-level debugging interface exists on the hardware as on the simulator, enabling the user to migrate from one environment to the another without making any changes to their design or their test-benches.

### 3.4. Scripting

While debugging large parallel systems, operations such as viewing the source code or variable values of individual processes become too low level; this is analogous to debugging a compiled process by inspecting its raw machine code and register values. For large parallel systems it is more convenient to be able to abstract the debugger to provide a higher, system-level interface. Such an interface allows the details of individual processes to be hidden, and replaced by system-specific displays. Clearly, it is impossible to provide interfaces for every possible system, so instead the debugger can be programmed using Tcl/Tk [7]. This allows the users to create their own system-specific interfaces, built on top of the debugger.

### 3.5. FileIO

When testing and debugging it is common to wish to use disk files in order to inject data into a system or to record intermediate results. This is achieved by providing an AE template which interfaces to the picoBus in the usual way using signals but which is also "connected" to a Unix file. The advantage of this method is that the same user's code can be used whether the system is running as a simulation or on hardware. The FileIO AE has two different implementations, one for simulation and one for hardware. In a simulation the connection to the file is simple, since the simulation simply consists of a piece of compiled C++. In hardware, the memory of the AE is used to buffer the data and when required the AE requests the debugger to either empty its memory (for an output FileIO) or fill it (for an input FileIO).

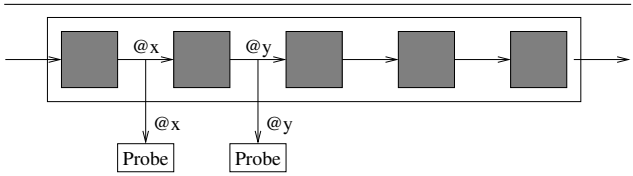

**Figure 4. Probe insertion**

### 3.6. Probes

Probes are special purpose processes inserted into the user's design by utilising unused processors. Probes can be connected to one or more signals, and can non-intrusively monitor all traffic which passes over the signals, as shown in figure 4. They do this by using the bus interconnect's ability to create 1-to-many connections. For example, suppose two processes in a system were connected by a 1-to-1 signal. If a probe is inserted during debugging to monitor that signal, the debug tools will change the 1-to-1 signal into a 1-to-many signal, with the probe acting as an extra destination. The original processes are unaffected by this change (both in terms of latency and bandwidth), but the probe is now able to monitor all communication over that signal.

Probes are implemented as processes, and so can run at full hardware speed. This enables probes to be used to debug systems in real-time. One use for probes is to allow real-time signal traces to be performed. Other uses include signal assertions, and on-chip analysis of data being processed.

Signal assertion probes can be used to check that the data passing over a signal conforms to some compile-time specified property. For example, all signals in picoArray devices have pre-allocated bandwidth. A signal assertion probe could be attached to a signal to record the bandwidth actually used, thus allowing signals with over-allocated bandwidth to be detected.

Probes can be used to perform on-chip analysis of signal data, rather than having to transport the data off-chip (e.g., using traces) for later analysis. For example, during the development of an in-house base station, a probe was created which performed Bit-Error Rate (BER) computation on signals. These BER probes could be used to monitor the performance of the base station's Viterbi decoder's in real-time, under different system loads.

## 4. A Method for Design and Debug

This section describes a typical process for creating a picoArray based application.

### 4.1. System Decomposition

Typically, this is done by hierarchically breaking down the problem into components consisting of processes connected by signals. Experience has shown that components generally contain a few tens of processes, however the number of processes required does not have to be specified at this stage. The boundaries of these components will also have signals defined and will eventually be connected to other parts of the system. Knowledge of the real-time system being developed is used to specify signal properties, such as maximum bandwidth and signal type. The properties can be checked during integration using signal assertions, which are described in section 4.3.

### 4.2. Component Coding

Two approaches can be taken, the choice being dictated by the complexity of the component.

For small components in which the division into AEs can be determined easily these AEs can be coded using C or assembler and connected using appropriate signals.

For larger components it may be preferable to initially produce a functional representation using C. This can be simulated even when the code size exceeds the memory for any AE, and allows functional testing of this component prior to its division into individual AEs.

Whichever approach is used the code can be tested by creating test harnesses using FileIO (see section 3.5) to mimic the external components. The symbolic debugger and its attendant tools can be used to find bugs within the AEs.

The migration of the code to hardware is eased by the fact that the same FileIO test harnesses produced for simulation can be used for verification. This highlights a huge advantage of this approach, since testing on hardware can be performed at a very early stage. This means that components can be tested for minutes or hours of real time, which would be impossible using simulation.

### 4.3. Small Scale Integration

As components are completed they can be integrated. The strong typing, bandwidth allocation, and fixed process creation ensure that components developed by different people will fit together properly. Signal assertions can be written to encode properties (such as signal value or minimum throughput) of the signals, and these can be checked during integration using assertion probes.

If integration fails (components fail to communicate properly), then this is caused by problems between components, rather than within a component (since the component has been verified in isolation, has static processes, fixed local signals, etc.). The suite of system-wide tools (probes, design browser, etc.) can be used to identify the problem.

### 4.4. Large Scale Integration and Testing

This phase of development can only really be done on the hardware. At this stage all of the FileIO will have been replaced by real components.

It is important to be able to monitor aspects of performance in real-time and this can be done using customised probes which monitor various signals and compare data throughput against predetermined limits. In addition, it is possible to monitor the behaviour of the system when processing real-world data, and to inject data by using the EPI. The results of the monitoring can be displayed using custom GUI's which the user can easily develop.

### 4.5. Comparison with Traditional Techniques

The power of the overall approach described here is that once a component has been written and tested, it can be assumed to work from that point on. Other parallel systems can behave like this for individual components, but then fail to work during integration or, even worse, during customer use. Possible integration problems include:

- priority inversion (e.g., Mars Pathfinder [6]).
- rogue processes corrupting shared memory
- overflowing message queues
- scheduling failures (e.g., improperly bounded or excessively large critical sections).
- multi-processor bus contention causing non-deterministic communication delays
- incoherency of multi-processor caches

Some or all of these problems will afflict other types of parallel systems, from multi-threaded programs containing just two processes through to large scale multi-processors. These problems are difficult to track down because they defy logical analysis, may behave non-deterministically, may only fail very infrequently, or may disappear when debug code is inserted to find the problem. Even if the cause of the problem is found, it is often difficult to write verification to prevent the bug recurring, because of the need to verify the entire system, not just the component or interface which causes the problem. In the worst case, verification may not be possible because it is not apparent why a fix actually works!

This solution avoids these types of problem in integration. Individual components which have been properly verified in isolation will behave in the same way when integrated into a complete system. The system behaves deterministically, so if problems are found, they can be reproduced, isolated, fixed, and verified. The overall development of systems is more predictable (timescales, etc.), since development isn't held up by strange problems which can't be found until integration.

## 5. Conclusion

In order to address its target markets in wireless communications, a family of code compatible devices has been created (currently PC101 and PC102) using the picoArray concept that provides the large computation power required by these applications. To ensure the largest possible computational resource, and recognising that system-wide debug is a major problem in multi-processor designs, a rethink of how to apply debugging was necessary. Therefore it was necessary to shift the debug burden from the hardware to the extensible software tools. The tools provide a way of debugging both single processors and more importantly the large multiprocessor systems possible on a picoArray, or indeed an array of picoArray devices. Experience has shown that using these tools it is possible to construct large working systems (currently in excess of 1000 processors).